\documentclass[epj]{webofc}
\usepackage[varg]{txfonts}
\usepackage{slashed}
\usepackage{subfigure}

\wocname{EPJ Web of Conferences}

\woctitle{INPC 2013}

\begin{document}

\title{Dense nucleonic matter and the renormalization group}

\author{Matthias Drews\inst{1,2}\fnsep\thanks{\email{matthias.drews@ph.tum.de}} \and
        Thomas Hell\inst{1,2} \and
        Bertram Klein\inst{1} \and
        Wolfram Weise\inst{1,2}
}

\institute{Physik Department T39, Technische Universit\"at M\"unchen, 85748 Garching, Germany
\and
           ECT*, Villa Tambosi, I-38123 Villazzano (Trento), Italy
          }

\abstract{
	Fluctuations are included in a chiral nucleon-meson model within the framework of the functional renormalization group. The model, with parameters fitted to reproduce the nuclear liquid-gas phase transition, is used to study the phase diagram of QCD. We find good agreement with results from chiral effective field theory. Moreover, the results show a separation of the chemical freeze-out line and chiral symmetry restoration at large baryon chemical potentials. 
}

\maketitle

\section{Introduction}
The quest for the critical end point of a first-order chiral transition in the QCD phase diagram is still unsettled. Lattice calculations at imaginary chemical potential seem to disfavor a first-order transition \cite{Focrand2007}. Model calculations are so far not conclusive. In the Polyakov-loop-extended quark-meson model, studied using the functional renormalization group, the critical endpoint lies in an unphysical region at very small temperatures and outside the range of applicability of the model \cite{Herbst2011}. In the Polyakov--Nambu--Jona-Lasinio model, the existence of the critical endpoint depends crucially on its input parameters such as the strengths of the axial anomaly and vector couplings \cite{PNJL}. \\
A related question is whether there exists a connection between chiral symmetry restoration and the chemical freeze-out line as determined in a hadron-resonance-gas model analysis of heavy-ion collision data \cite{Andronic2009}. For very small baryon chemical potentials the chiral crossover lies close to the chemical freeze-out points, not unexpectedly since chemical equilibiration requires multi-particle effects and collective phenomena right at the borderline of hadronization mostly into pions \cite{BraunMunzinger2004}. At larger baryon chemical potentials it is mandatory to take all the well-established constraints from nuclear physics into account. A way to do this is to start with a model based on the relevant nucleonic and mesonic degrees of freedom. In order to study chiral restoration, a chiral nucleon-meson model is chosen \cite{Berges2003}. No relationship between chiral restoration and chemical freeze-out points is found in this model, at least not in the mean-field approximation \cite{Floerchinger2012}. \\

In the following, we will briefly review the chiral nucleon-meson model and its mean-field treatment. Then we strengthen the conclusions by providing a self-consistent treatment of thermal mesonic and nucleonic fluctuations in the framework of the functional renormalization group (FRG). The phase diagram around the nuclear liquid-gas transition is compared with results of calculations using chiral effective field theory \cite{Fiorilla2012}.

\section{The chiral nucleon-meson model}
The dominant degrees of freedom around the nuclear liquid-gas transition are nucleons and pions. The pions are combined with a scalar field, $\sigma$, into a four-component vector, $\phi=(\sigma, \boldsymbol \pi)$\,, that transforms under $\operatorname{SO}(4)\cong \operatorname{SU}(2)\times\operatorname{SU}(2)$, with the invariant $\rho = {1\over 2} |\phi|^2 = {1\over 2} (\sigma^2 + \boldsymbol \pi\cdot \boldsymbol \pi)$. Moreover, the nucleon is coupled to an isoscalar vector field, $\omega_\mu$, generating a repulsive short-range nucleon-nucleon interaction. The Lagrangian of the chiral nucleon-meson model reads \cite{Berges2003}:
\begin{align*}
	\mathcal L&=\bar\psi\Big[i\slashed \partial+g_s(\sigma+i\gamma_5\boldsymbol\pi\cdot\boldsymbol\tau)-g_v\gamma^\mu\omega_\mu+\gamma^0\mu\Big]\psi+\partial_{[\mu}\omega_{\nu]}\partial^{[\mu}\omega^{\nu]} +\tfrac 12m_v^2\omega_\mu\omega^\mu+\\
	&\quad+\tfrac 12\partial_\mu\sigma\,\partial^\mu\sigma+\tfrac 12\partial_\mu\boldsymbol\pi\cdot\partial^\mu\boldsymbol\pi+U_{\text{mic}}(\rho,\sigma)\,.
\end{align*}
While the microscopic potential, $U_{\text{mic}}$, is unknown a priori, the relevant object of interest is the effective potential $U$ at a given temperature and baryon chemical potential with respect to the potential right at the equilibrium point of nuclear matter, $U(T,\mu)-U(T=0,\mu=\mu_c)\,.$  Here, $\mu_c=M_N - B = 923\operatorname{ MeV}$, i.\,e., the difference of the nucleon mass and the binding energy per nucleon, coincides with the critical chemical potential at the $T = 0$ intercept of the liquid-gas phase transition line. In the mean-field approximation the spatial components of the $\omega_\mu$ field vanish in order to preserve rotational invariance. In addition, we assume that there is no pion condensate, so only the mean-field values of the $\sigma$ and the $\omega_0$ will contribute. The nucleons can be integrated out and the effective potential takes the form
\begin{align*}
	U_{\text{MF}}=U(\sigma,\omega_0)-4T\int\frac{d^3p}{(2\pi)^3}\log\bigg[1+\operatorname{e}^{-\big(E_N(p)-\mu_{\text{eff}}\big)/T}\bigg]-4T\int\frac{d^3p}{(2\pi)^3}\log\bigg[1+\operatorname{e}^{-\big(E_N(p)+\mu_{\text{eff}}\big)/T}\bigg]\,,
\end{align*}
with $E_N(p)^2=p^2+(g_s\sigma)^2$ and $\mu_{\text{eff}}=\mu-g_v\omega_0$. The potential $U(\sigma,\omega_0)$ is chosen in such a way as to reproduce nuclear physics constraints \cite{Floerchinger2012}. The mean-field potential $U_{\text{MF}}$ is then minimized as a function of $\sigma$ and $\omega_0$. Next we explain how to go beyond this mean field approximation.

\section{Adding fluctuations}
In order to incorporate fluctuations in a self-consistent manner, the functional renormalization group (FRG) method is used (\cite{Kopietz2010} and references therein). The effective action, $\Gamma_k$, at a renormalization scale $k$, interpolates in this framework between the microscopic action, $S=\Gamma_{k=\Lambda}$, at a cutoff scale $\Lambda$ and the full effective action, $\Gamma_{\text{eff}}=\Gamma_{k=0}$, with all fluctuations integrated out. The flow of this action, as the scale $k$ is lowered, is governed by Wetterich's flow equation \cite{Wetterich1993}:
\begin{align*}
	k\partial_k\Gamma_k=
	\begin{aligned}
		\vspace{1cm}
		\includegraphics[width=0.1\textwidth]{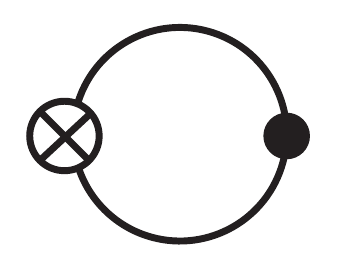}
	\end{aligned} \vspace{-1cm} = \frac 12 \operatorname{Tr}\frac{k\partial_kR_k}{\Gamma^{(2)}+R_k}\,.
\end{align*}
The trace is taken over all fields, as well as their momenta and internal degrees of freedom. $\Gamma^{(2)}$ is the second derivative of the effective action with respect to the fields, such that the last expression is the full propagator, with an insertion of a regulator function $R_k$. This regulator function ensures that only fluctuations around $k$ are contributing at that scale.  The regulator gives low momentum modes an effective mass of order $k^2$. In practice, $R_k$ is chosen to be the regulator proposed by Litim \cite{Litim2001} for application at finite temperatures \cite{Litim2006,Blaizot2006}:
\begin{gather*}
	R_k =(k^2-\boldsymbol p^2)\cdot\theta(k^2-\boldsymbol p^2)\,.
\end{gather*}
The potential is fitted to reproduce nuclear physics constraints such as the liquid-gas transition. The effective potential is therefore expanded around $T=0$ and $\mu=\mu_c$ and the flow of
\[
	\bar \Gamma_k=\Gamma_k(T,\mu)-\Gamma_k(0,\mu_c)
\]
is computed. Thermal fluctuations around the liquid-gas transition are treated self-consistently in this way \cite{Litim2006}. In the local-potential approximation and in leading order in the derivative expansion, the flow equation for the effective action, $\bar\Gamma_k$, reduces to an equation for the effective potential, $\bar U_k$. The trace can be computed explicitly and the flow equation for the effective potential becomes
\begin{gather*}
	\partial_k\bar U_k=f(T,\mu)-f(0,\mu_c)\,,
\end{gather*}
where
\begin{gather*}
	f(T,\mu)=\frac {k^4}{12\pi^2} \bigg\{\frac{3\big[1+2n_B(E_\pi)\big]}{E_\pi}+\frac {1+2n_B(E_\sigma)}{E_\sigma}-\frac{8\big[1-n_F(E_N,\mu_{\text{eff}})-n_F(E_N,-\mu_{\text{eff}})\big]}{E_N}\bigg\}\,.
\end{gather*}
Here
\begin{gather*}
	E_\pi^2=k^2+m_\pi^2\,,\quad E_\sigma^2=k^2+m_\sigma^2\,, \quad E_N^2=k^2+2g_s^2\,\rho\,, \quad m_\pi^2=U_k'(\rho)\,, \quad m_\sigma^2=U_k'(\rho)+2\rho U_k''(\rho)\,, \\
	\quad m_N=g_s\sigma\,,\quad \mu_{\text{eff}}=\mu-g_v\,\omega_{0,k}\,,\quad n_B(E)=\frac 1{\operatorname{e}^{\beta E}-1}\,\text{~ and ~}  n_F(E,\mu)=\frac 1{\operatorname{e}^{\beta(E-\mu)}+1}\,.
\end{gather*}
The flow equation for the background vector field $\omega_{0,k}$ leads to the integral equation
\[
\omega_{0,k}={2g_v\over 3\pi^2\,m_v^2}\int_k^\Lambda d\tilde k\,{\tilde k}^4\;\frac{\partial}{\partial\mu}\frac{n_F\big(E_N(\tilde k),\mu_{\text{eff}}\big)+n_F\big(E_N(\tilde k),-\mu_{\text{eff}}\big)}{E_N(\tilde k)}\,.
\]
The effective potential is fixed in such a way that the nuclear saturation density, the binding energy, the surface tension of a nuclear droplet, and the compression modulus obtained from the full potential agree with empirical data. The flow equation is then solved for temperatures and baryon chemical potentials around the liquid-gas phase transition using a numerical method on a discretized grid \cite{Adams1995}.

\section{Results and discussion}
\begin{figure}
	\centering
	\includegraphics[width=0.48\textwidth]{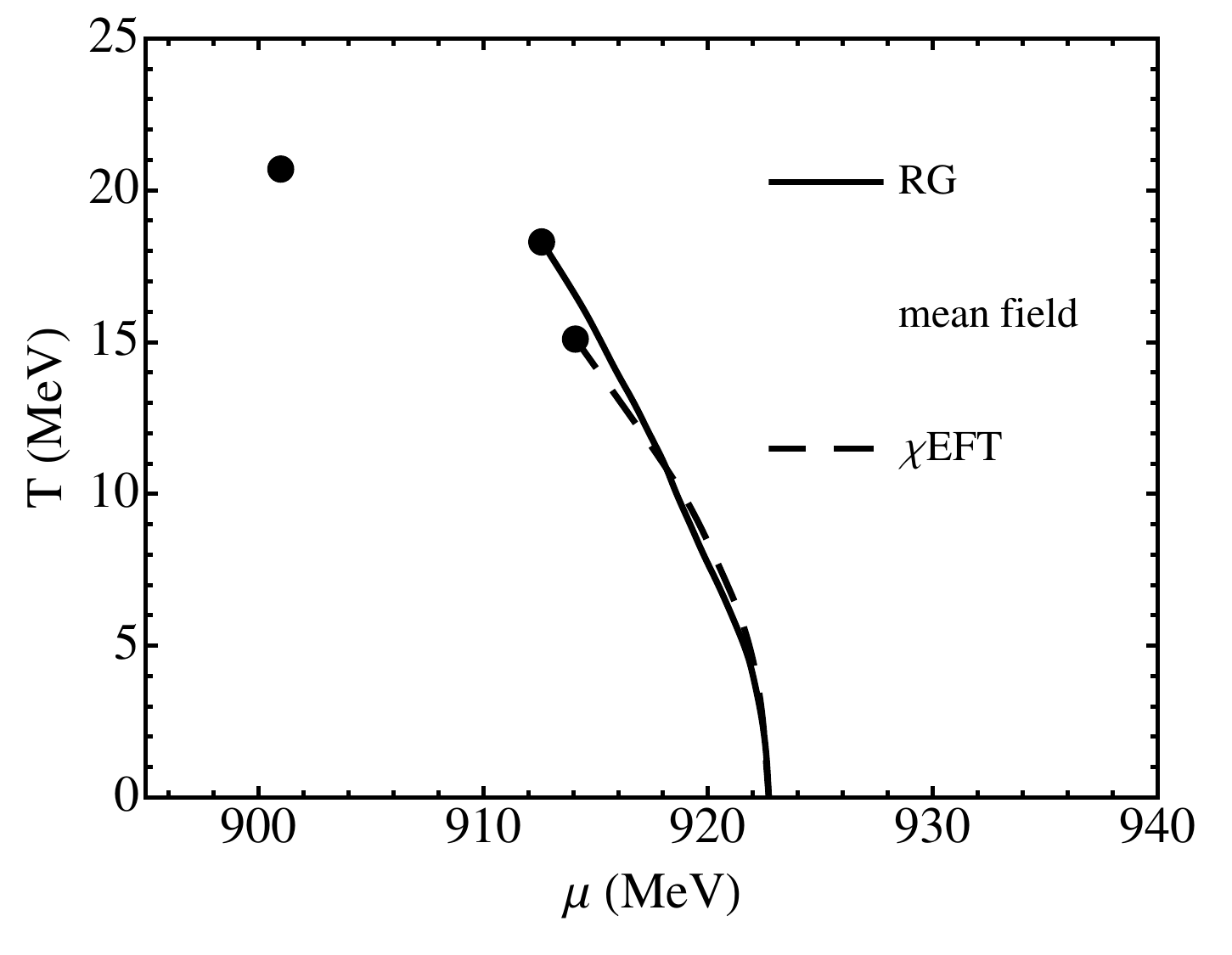}\qquad
	\includegraphics[width=0.47\textwidth]{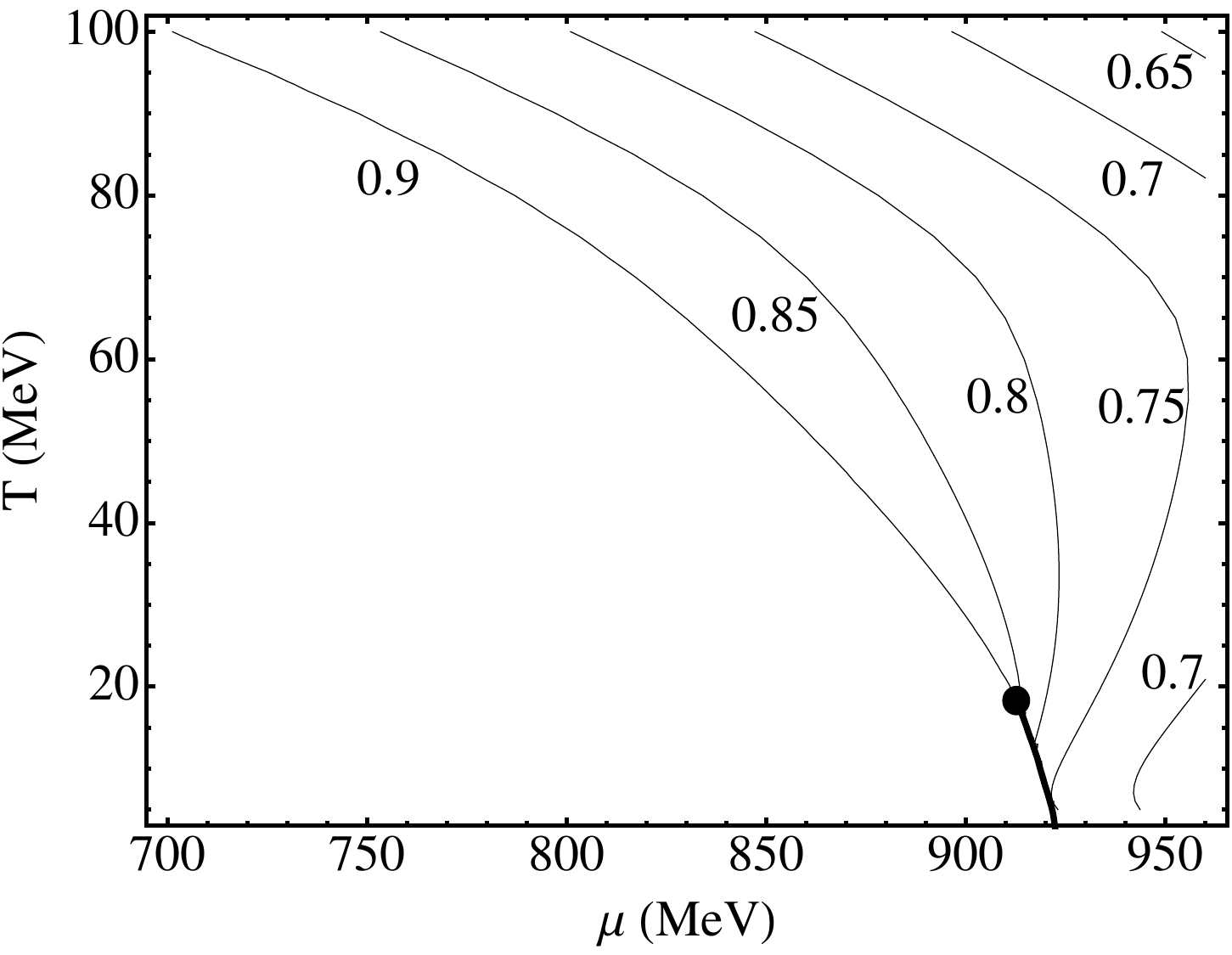}
	\caption{Left: liquid-gas phase transition in $\chi$EFT \cite{Fiorilla2012} (dashed) and in the chiral nucleon-meson model both at mean-field level (dotted) and with fluctuations (solid). Right: contour plots of $\sigma/f_\pi$ representing the chiral order parameter. Within the region of applicability, the condensate is non-zero throughout and chiral symmetry is spontaneously broken.\label{fig}}
\end{figure}
The nuclear liquid-gas phase transition has been studied extensively within the framework of chiral effective field theory ($\chi$EFT, \cite{Holt2013} and references therein). In the left plot of Fig.~\ref{fig} the first-order line and the critical point are shown, as computed in the nucleon-meson model, both at mean-field level and including fluctuations. They are compared with a $\chi$EFT computation \cite{Fiorilla2012} which takes into account all one- and two-pion exchange processes, as well as three-body forces and $\Delta$-isobar excitations. With fluctuations, the transition is bent away from the mean-field curve to higher chemical potentials, in very good agreement with $\chi$EFT results. This is remarkable, given the very different approaches. Whereas the temperature of the critical point is $15.1\operatorname{ MeV}$ in $\chi$EFT, it is shifted to a value of $T_c=18.3\operatorname{ MeV}$ in the RG treatment of the nucleon-meson model, in good agreement with empirical results \cite{Karnaukhov2008}. 

In order to address the entanglement between chemical freeze-out points and chiral restoration, the chiral condensate is studied as a function of temperature and chemical potential. In the nucleon-meson model, the chiral condensate is proportional to the expectation value of $\sigma$. In the plot on the right-hand side of Fig.~\ref{fig}, contour lines of $\sigma$ normalized to its vacuum expectation value, $f_\pi$, are shown. In the whole area of temperatures up to $100$~MeV and baryon chemical potentials smaller than about $1$~GeV, the chiral order parameter $\sigma/f_\pi$ still exceeds $0.65$ and chiral symmetry is not restored in its trivial Wigner-Weyl realization. The line at which chiral symmetry is restored must therefore intersect the $\mu$-axis at considerably larger baryon chemical potentials and is therefore well separated from the nuclear liquid-gas phase transition, as it should be. In the region of applicability, there is no sign of a chiral first-order phase transition. This demonstrates the importance of taking the constraints from nuclear physics properly into account in calculations modeling the QCD phase diagram.

\end{document}